%% file: main.tex
\documentclass[10pt,conference]{IEEEtran}
\IEEEoverridecommandlockouts

\usepackage{cite}
\usepackage{amsmath,amssymb,amsfonts}
\usepackage{textcomp}
\usepackage{xcolor}
\usepackage[most]{tcolorbox}
\usepackage{balance}
\usepackage{enumitem}
\usepackage{hyperref}
\usepackage{listings} 
\usepackage{xcolor}
\usepackage{microtype} 
\usepackage{xspace}
\usepackage{algorithm}
\usepackage{algpseudocode}
\usepackage{stmaryrd}
\usepackage{soul}
\usepackage{color, colortbl}
\usepackage{amsfonts} 
\usepackage{booktabs} 
\usepackage{graphicx, subfigure}
\usepackage{epigraph} 
\usepackage{framed}
\usepackage{here}
\definecolor{coolblack}{rgb}{0.0, 0.18, 0.39}
\usepackage{multirow}
\usepackage{xurl}
\usepackage{comment}
\usepackage{soul}
\usepackage{here}

\definecolor{awesome}{rgb}{0.0, 0.2, 0.6}
\input{comments}

\def\BibTeX{{\rm B\kern-.05em{\sc i\kern-.025em b}\kern-.08em
    T\kern-.1667em\lower.7ex\hbox{E}\kern-.125emX}}
    
\begin{document}

\title{Understanding the Characteristics of LLM-Generated Property-Based Tests in Exploring Edge Cases}


\author{\IEEEauthorblockN{Hidetake Tanaka}
\IEEEauthorblockA{\textit{Nara Institute of Science and Technology} \\
Ikoma, Japan \\
tanaka.hidetake.te0@naist.ac.jp}
\and
\IEEEauthorblockN{Haruto Tanaka}
\IEEEauthorblockA{\textit{Nara Institute of Science and Technology} \\
Ikoma, Japan \\
tanaka.haruto.td7@is.naist.jp}
\and
\IEEEauthorblockN{Kazumasa Shimari}
\IEEEauthorblockA{\textit{Nara Institute of Science and Technology} \\
Ikoma, Japan \\
k.shimari@is.naist.jp}
\and
\IEEEauthorblockN{Kenichi Matsumoto}
\IEEEauthorblockA{\textit{Nara Institute of Science and Technology} \\
Ikoma, Japan \\
matumoto@is.naist.jp}
}

\maketitle
\begin{abstract}
As Large Language Models (LLMs) increasingly generate code in software development, ensuring the quality of LLM-generated code has become important. Traditional testing approaches using Example-based Testing (EBT) often miss edge cases—defects that occur at boundary values, special input patterns, or extreme conditions. This research investigates the characteristics of LLM-generated Property-based Testing (PBT) compared to EBT for exploring edge cases. We analyze 16 HumanEval problems where standard solutions failed on extended test cases, generating both PBT and EBT test codes using Claude-4-sonnet. Our experimental results reveal that while each method individually achieved a 68.75\% bug detection rate, combining both approaches improved detection to 81.25\%. The analysis demonstrates complementary characteristics: PBT effectively detects performance issues and edge cases through extensive input space exploration, while EBT effectively detects specific boundary conditions and special patterns. These findings suggest that a hybrid approach leveraging both testing methods can  improve the reliability of LLM-generated code, providing guidance for test generation strategies in LLM-based code generation.
\end{abstract}


\input{contents/section1}
\input{contents/section2}
\input{contents/section3}
\input{contents/section4}

\input{contents/section5}
\input{contents/section6}

\section*{Acknowledgment}
This work has been supported by JSPS KAKENHI Number JP23K16862, and JST BOOST, Japan Grant Number JPMJBS2423.

\bibliographystyle{IEEEtranS.bst}
\bibliography{reference}

\end{document}

%% file: comments.tex
\usepackage{datenumber}

\definecolor{chestnut}{rgb}{0.8, 0.36, 0.36}

\definecolor{chestnut}{rgb}{0.8, 0.36, 0.36}

\newcounter{dateone}\newcounter{datetwo}%
\newcommand{\daydifftoday}[3]{%
\setmydatenumber{dateone}{\the\year}{\the\month}{\the\day}%
\setmydatenumber{datetwo}{#1}{#2}{#3}%
\addtocounter{datetwo}{-\thedateone}%
\thedatetwo
}

%% file: contents/section1.tex
\section{Introduction}


In recent years, the development of Large Language Models (LLMs) has led to rapid advances in technologies that automatically generate program code from natural language instructions. LLMs such as Claude\footnote{https://claude.ai/} can generate high-quality code even for complex programming tasks, significantly contributing to the efficiency of software development. Furthermore, approaches that iteratively improve code by generating code using LLMs and providing feedback on code improvements from developers to LLMs are becoming increasingly common.


However, code generated by LLMs still faces reliability challenges. In particular, it is difficult to automatically test whether the generated code meets specifications. Current mainstream approaches aim to improve reliability by having LLMs iteratively perform code generation and test execution, but most of these rely on example-based unit testing~\cite{unittest_survey}. Figure \ref{fig:ebt-example} shows the Example-based Testing (EBT) approach using a function that returns the maximum value from a list as an example. This approach only executes tests on limited input-output examples, potentially missing bugs for inputs not covered by test cases. These bugs are often caused by edge cases, such as boundary values, special input patterns, or extreme conditions.

\input{figures/ebt-example}


Property-based Testing (PBT)~\cite{claessen2000quickcheck} is a method that defines general properties that code should satisfy and tests whether these properties are maintained against automatically generated diverse input data. Figure \ref{fig:pbt-example} illustrates the PBT approach for the same maximum value function. By using this approach, it becomes possible to automatically cover a wide range of input space for the defined properties.

\input{figures/pbt-example}


This research aims to reveal practical insights into the effectiveness of Property-based Testing (PBT) and Example-based Testing (EBT) for exploring edge cases in LLM-based code generation. Through comparative evaluation of both methods, we quantitatively analyze the characteristics and limitations of each approach to gain insights into effective testing strategies for LLM-based test generation. The obtained insights contribute to improving the quality of LLM-generated code and providing guidelines for test method selection, helping establish effective testing strategies in real development environments.


The structure of this paper is as follows. First, Section \ref{sec:related} describes related work, and Section \ref{sec:case-study-design} explains the case study design. Next, Section \ref{sec:results} reports the experimental results, and Section \ref{sec:discussion} provides discussion. Finally, Section \ref{sec:conclusion} summarizes this research.

For reproducibility of the case study, the replication package is available on Zenodo\footnote{https://doi.org/10.5281/zenodo.16434613}.

%% file: figures/ebt-example.tex
\begin{figure}[!t]
\begin{lstlisting}[
    basicstyle=\small
]
def test_biggest():
    assert biggest([1, 2, 3]) == 3
    assert biggest([3, 2, 1]) == 3
    assert biggest([-1, -5, -3]) == -1
    assert biggest([42]) == 42
\end{lstlisting}
\caption{Example-based Testing approach}
\label{fig:ebt-example}
\end{figure}

%% file: figures/pbt-example.tex
\begin{figure}[!t]
\begin{lstlisting}[
    basicstyle=\small
]
@given(st.lists(st.integers(), min_size=1))
def test_biggest(lst):
    assert biggest(lst) == sorted(lst)[-1]
\end{lstlisting}
\caption{Property-based Testing approach}
\label{fig:pbt-example}
\end{figure}

%% file: contents/section2.tex
\section{Related Work}
\label{sec:related}



\subsection{LLM-based Code Generation}

LLMs have been applied to various tasks in software engineering, and research on their utilization has become increasingly active. A systematic literature review by Hou et al.~\cite{hou2024large} reported that approximately 70\% of LLM research in software engineering targeted generation tasks such as code generation, test generation, and documentation generation.


Research on code generation using LLMs had been rapidly developing in recent years. According to a comprehensive survey by Jiang et al.~\cite{jiang2024survey}, this field had made progress across diverse aspects, from functional correctness improvement to code efficiency, security, and maintainability. Chen et al.~\cite{chen2021evaluating} investigated the performance of generating Python code from natural language descriptions using OpenAI's Codex model\footnote{https://openai.com/index/openai-codex/}, showing the potential of LLM-based code generation. Subsequently, commercial tools such as GitHub Copilot\footnote{https://github.com/features/copilot} emerged one after another, and their utilization in development environments advanced rapidly. Google reported that AI generated more than a quarter of new code~\cite{google2024}, demonstrating the widespread adoption of LLM-based code generation.


Research on improving the quality of code generated by LLMs had also been actively conducted, with particular attention to multi-agent approaches. Huang et al.~\cite{huang2024agentcoder} proposed AgentCoder, a multi-agent code generation framework. AgentCoder employed a mechanism where multiple agents responsible for code generation, test design, and test execution worked collaboratively, using test results as feedback to iteratively improve code. When this method was evaluated using GPT-4, it achieved pass@1 rates (probability of correct answer on the first attempt) of 96.3\% and 91.8\% on the HumanEval and MBPP datasets, respectively. Furthermore, Islam et al.~\cite{islam2024mapcoder} proposed MapCoder for competitive programming, which employed four agents (recalling, planning, code generation, and debugging) that mimicked the program synthesis cycle performed by human developers, achieving 93.9\% pass@1 on HumanEval. Additionally, Dong et al.~\cite{dong2024self} adopted a similar approach using ChatGPT, where multiple agents responsible for code generation, test design, and test execution worked collaboratively, utilizing test results as feedback to iteratively improve code. These results demonstrated significant performance improvements compared to conventional methods, proving the effectiveness of the multi-agent approach.


To support more accurate code generation, research had also been conducted on applying tests to code generated by LLMs. Fakhoury et al.~\cite{fakhoury2024llm} proposed TiCoder, an interactive method that resolved ambiguity when generating code from natural language using test cases. In this method, LLMs generated multiple test cases along with code candidates, and users selected test cases that matched their intentions, thereby clarifying user intent while narrowing down the code. Benchmark evaluations showed that this method improved pass@1 by an average of 45.97\%.


Despite these high success rates, code generated by LLMs still often contained edge case bugs. Even code that showed high performance on standard benchmarks often behaved unexpectedly under boundary conditions or special input patterns. This issue was thought to stem from the gap between common patterns included in LLM training data and the diverse edge cases encountered in actual software development.


\subsection{Property-based Testing}


Property-based Testing (PBT) is a method originating from QuickCheck, proposed by Claessen et al.~\cite{claessen2000quickcheck}. This method defines properties that code should satisfy and tests whether these properties are maintained against automatically generated diverse input data. By using this approach, it becomes possible to automatically cover a wide range of input space for the defined properties. Hatfield-Dodds~\cite{hatfield2020falsify} demonstrated how PBT was used to discover bugs in existing libraries, showing the importance of introducing PBT.


There are several testing methods that share conceptual similarities with PBT, such as Fuzzing~\cite{fuzzing_survey} and Mutation Testing~\cite{mutation_testing}. Fuzzing generates random or semi-random inputs to detect abnormal program behavior, exploring a wide input space similar to PBT. Mutation Testing intentionally introduces changes (mutants) to the program to evaluate test quality, but is limited to assessing existing tests. On the other hand, PBT explicitly defines expected behavior as properties, enabling more structured and comprehensive bug detection.


In recent years, research had begun on applying PBT to code generated by LLMs. Vikram et al.~\cite{vikram2024llm} researched whether PBT test code could be automatically generated based on API documentation. More recently, He et al.~\cite{he2025pbt} proposed the ``Property-Generated Solver'' framework, which used PBT to validate LLM-generated code, achieving relative improvements of 23.1\% to 37.3\% in pass@1 compared to traditional Test-Driven Development (TDD) methods. These studies have demonstrated the feasibility and effectiveness of LLM-based PBT generation. However, the characteristics of PBT test code generated by LLMs had not yet been sufficiently studied in the context of edge case detection.


Based on these insights, this research clarifies the characteristics of PBT for edge case exploration through comparative analysis of PBT and EBT in LLM-based test generation, which has not been sufficiently examined in existing studies. Focusing particularly on the intrinsic bug detection capabilities of PBT demonstrated by Hatfield-Dodds, we clarify the characteristics these methods exhibit in the context of LLM-generated test code. These insights provide guidance for achieving more reliable code generation in LLM-powered software development.

%% file: contents/section3.tex
\section{Case Study Design}
\label{sec:case-study-design}


This section describes the design of a case study that comparatively evaluates Property-based Testing (PBT) and Example-based Testing (EBT) for their edge case detection capabilities in LLM-generated code. He et al.~\cite{he2025pbt} had proposed a PBT-based validation framework for LLM-generated code and demonstrated improvements in pass@1. To clarify the characteristics of PBT, systematic comparison with the traditional mainstream approach of EBT is essential. EBT is the commonly used testing method in current software development, as shown by the survey by Daka et al.~\cite{unittest_survey}. This research quantitatively analyzes the characteristics of both methods specifically from the perspective of edge case detection, clarifying the unique strengths and limitations of each approach.


\subsection{Research Questions}


This case study aims to answer the following two research questions:


\textbf{RQ1: How effective are PBT and EBT respectively in detecting edge cases in LLM-generated code?}


This research question quantitatively evaluates how many bugs each method can detect across 16 test cases containing edge cases. By comparing detection rates, we clarify the fundamental effectiveness of each approach.


\textbf{RQ2: What types of edge cases do each method fail to detect, and what are the underlying causes?}


This research question analyzes in detail the edge cases that each method fails to detect and classifies failure patterns. This identifies the limitations of each approach and areas for improvement.


\subsection{Experimental Design}


This case study employs two datasets: the HumanEval dataset~\cite{chen2021evaluating} for evaluating LLM code generation capabilities, and the HumanEval+ dataset~\cite{evalplus}, which extends HumanEval's test cases by approximately 80-fold.


HumanEval is a representative dataset for evaluating the code generation capabilities of large language models (LLMs), consisting of 164 programming problems. Each problem defines a task of generating functionally correct Python code based on a natural language specification given as a docstring. The problems cover a wide range from basic string operations, numerical calculations, and list processing to algorithm implementations, enabling comprehensive evaluation of fundamental programming skills. Additionally, each problem comes with an average of 7.7 unit test cases, which serve as criteria for evaluating the correctness of the generated code.


HumanEval+ is a dataset that extends the test cases of HumanEval by approximately 80 times. It had been proven through extensive evaluations on 26 popular LLMs (e.g., GPT-4 and ChatGPT) that HumanEval+ could detect a significant amount of previously undetected wrong code synthesized by LLMs, reducing the pass@k by up to 19.3-28.9\%.


To focus the evaluation on edge case detection, we construct the analysis set by starting from HumanEval solutions that failed under HumanEval+ and filtering for issues related to boundary conditions, performance, or input structures.


We first apply the HumanEval+ extended test cases to all 164 standard solutions provided by HumanEval, and 21 of the 164 failed under these extended tests. Among these 21 failed cases, we then select 16 cases that are appropriate for evaluating edge case handling. The selection criteria are as follows: (1) bugs related to boundary conditions such as 0, negative numbers, or maximum/minimum values; (2) bugs related to performance issues or timeouts; or (3) bugs related to specific input structures or combinations. The remaining 5 failed cases do not fall into any of the above categories and are excluded because their function specifications are ambiguous, making it difficult to determine edge case bugs.


\subsection{Test Generation Methodology}


In this research, we generate both PBT and EBT test code for each problem using Claude-4-sonnet. To ensure accurate evaluation of edge case detection, it is necessary to eliminate false failures caused by defective test code itself. Test codes are excluded based on the following criteria: (1) those containing syntax errors or runtime errors, (2) PBT cases defining properties that do not conform to the function specifications, and (3) EBT cases with incorrect expected output values. This exclusion process enables clear distinction between test failures caused by edge case detection and those caused by test code defects.


\subsubsection{Property-based Testing (PBT) Generation}

PBT is a testing approach that defines properties that functions should satisfy and tests whether these properties are maintained with automatically generated diverse input data.


For PBT test code generation, we specified the use of Hypothesis~\cite{hypothesis}, Python's standard PBT library. In prompts, we clearly defined the role as a tester and provided detailed instructions including how to utilize the Hypothesis library. The prompts included the following elements:
\begin{itemize}
    \item \textbf{Role Definition}: Definition of the role as a tester responsible for creating property-based test cases
    \item \textbf{Detailed Instructions}: Comprehensive test implementation requirements including Hypothesis library usage, appropriate strategies selection, focus on properties and invariants, and handling of edge cases and large-scale inputs
    \item \textbf{Test Format Specification}: Standard PBT test structure using the \texttt{@given} decorator
    \item \textbf{Few-shot Examples}: Example prompts and completions from HumanEval dataset (abbreviated in figure) to guide the LLM's test generation
    \item \textbf{Input Code Snippet}: Target function code snippet from HumanEval dataset
\end{itemize}

\input{figures/pbt-prompt}


Figure \ref{fig:pbt-prompt} shows the specific prompt structure used for PBT test code generation. This prompt clearly defines the role as a tester and provides detailed instructions including how to utilize the Hypothesis library. Furthermore, it includes examples of specific prompts and expected outputs to leverage few-shot learning.


LLMs that receive the prompt in Figure \ref{fig:pbt-prompt} generate test code with the following universal properties based on function specifications:
\begin{itemize}
    \item \textbf{Threshold Boundary Conditions}: The condition is not satisfied at exactly the threshold distance, but only satisfied for distances less than the threshold
    \item \textbf{Empty List Behavior}: Always returns False for empty lists
    \item \textbf{Single Element Behavior}: Always returns False for single-element lists
\end{itemize}


The LLM typically generates 3-5 property-based tests, with each test expressed as an invariant that should hold across the entire input space. These properties do not directly specify expected output for specific input. Instead, PBT describes properties as universal constraints and tests function behavior by automatically generating diverse input data. This approach automatically executes tests that cover boundary values and edge cases that developers might not easily anticipate, effectively improving code robustness.


\subsubsection{Example-based Testing (EBT) Generation}

EBT is a testing approach that verifies code behavior using pairs of specific input values and their expected output values.


For EBT test code generation, we provided the LLM with prompts containing the following elements:
\begin{itemize}
    \item \textbf{Role Definition}: Definition of the role as a tester responsible for creating comprehensive test cases
    \item \textbf{Three Test Categories}: Basic Test Cases (verifying fundamental functionality under normal conditions), Edge Test Cases (evaluating behavior under extreme conditions), and Large Scale Test Cases (assessing performance and scalability with large data samples)
    \item \textbf{Detailed Instructions}: Comprehensive test implementation requirements, documentation specifications, emphasis on edge cases, and focus on performance testing
    \item \textbf{Test Format Specification}: Specification of test structure using assert statements
    \item \textbf{Few-shot Examples}: Example prompts and completions from HumanEval dataset (abbreviated in figure) to guide the LLM's test generation
    \item \textbf{Input Code Snippet}: Target function code snippet from HumanEval dataset
\end{itemize}


The LLM typically generates 10-20 specific test cases, with each test case expressed as a pair of specific input values and their expected output values. The characteristic of EBT lies in verifying function behavior through concrete examples that are intuitively understandable to developers. Particularly for boundary conditions and special cases, the LLM explicitly generates important edge cases inferred from function specifications as test cases.

\input{figures/ebt-prompt}


Figure \ref{fig:ebt-prompt} shows the specific prompt structure used for EBT test code generation. This prompt is based on the prompt used in the AgentCoder framework by Huang et al.~\cite{huang2024agentcoder}. The prompt clearly defines the role as a tester and instructs the generation of comprehensive test cases divided into three categories: Basic, Edge, and Large Scale.


\subsection{Evaluation Metrics}


This research measures the effectiveness of each method using the following evaluation metrics:


\subsubsection{Primary Metrics}


\textbf{Detection Rate}: The proportion of cases where each method successfully detected bugs out of 16 test cases. The criteria for successful bug detection are defined as the occurrence of either assertion failures or timeouts (15 seconds) during test execution.


\textbf{Failure Patterns}: Classification and analysis of edge cases where each method fails to detect bugs. Failure patterns are classified into the following categories:
\begin{itemize}
    \item Boundary-related: Boundary conditions such as 0, negative numbers, maximum/minimum values
    \item Performance-related: Timeout or efficiency issues with large-scale inputs
    \item Special patterns: Issues related to specific input structures or combinations
\end{itemize}


\subsubsection{Supplementary Metric}


\textbf{Execution Time}: Average time required for test execution by each method. While this metric is not a primary evaluation criterion, it is recorded as reference information from a practical perspective.


For the target code sets, we execute tests using the following two types of test code and compare their results:
\begin{enumerate}
    \item Property-based test code generated by LLM (Claude-4-sonnet)
    \item Example-based test code generated by LLM (Claude-4-sonnet)
\end{enumerate}

%% file: figures/pbt-prompt.tex
\begin{figure}[!t]
\begin{lstlisting}[
    basicstyle=\small, %or \small or \footnotesize etc.
]
**Role**: As a tester, create property-based test cases for the given function using Hypothesis to automatically generate diverse inputs and verify function properties.

**Instructions**:
- Use Hypothesis library with appropriate strategies for input generation
- Focus on properties and invariants the function should satisfy
- Document each test property with clear comments
- Handle edge cases and large-scale inputs

- The format of test cases should be:
```python
from hypothesis import given, strategies as st

@given(strategy_for_input)
def test_property_name(input_params):
    result = function_name(input_params)
    assert property_condition, "Property description"
```

# For example:

## Prompt 1:
...

## Completion 1:
...

## Prompt 2:
...

## Completion 2:
...

**Input Code Snippet**
```python
from typing import List

def has_close_elements(numbers: List[float], threshold: float) -> bool:
    """ Check if in given list of numbers, are any two numbers closer to each other than
    given threshold.
    >>> has_close_elements([1.0, 2.0, 3.0], 0.5)
    False
    >>> has_close_elements([1.0, 2.8, 3.0, 4.0, 5.0, 2.0], 0.3)
    True
    """
```
\end{lstlisting}
\caption{PBT test code generation prompt example}
\label{fig:pbt-prompt}
\end{figure}

%% file: figures/ebt-prompt.tex
\begin{figure}[!t]
\begin{lstlisting}[
    basicstyle=\small, %or \small or \footnotesize etc.
]
**Role**: As a tester, your task is to create comprehensive test cases for the incomplete function. These test cases should encompass Basic, Edge, and Large Scale scenarios to ensure the code's robustness, reliability, and scalability.

**1. Basic Test Cases**:
- **Objective**: To verify the fundamental functionality of the `has_close_elements` function under normal conditions.

**2. Edge Test Cases**:
- **Objective**: To evaluate the function's behavior under extreme or unusual conditions.

**3. Large Scale Test Cases**:
- **Objective**: To assess the function's performance and scalability with large data samples.

**Instructions**:
- Implement a comprehensive set of test cases following the guidelines above.
- Ensure each test case is well-documented with comments explaining the scenario it covers.
- Pay special attention to edge cases as they often reveal hidden bugs.
- For large-scale tests, focus on the function's efficiency and performance under heavy loads.

- The format of test cases should be:
```python
assert function_name(input) == expected_output, "Test Case Description"
```

# For example:

## Prompt 1:
...

## Completion 1:
...

## Prompt 2:
...

## Completion 2:
...

**Input Code Snippet**
```python
from typing import List

def has_close_elements(numbers: List[float], threshold: float) -> bool:
    """ Check if in given list of numbers, are any two numbers closer to each other than
    given threshold.
    >>> has_close_elements([1.0, 2.0, 3.0], 0.5)
    False
    >>> has_close_elements([1.0, 2.8, 3.0, 4.0, 5.0, 2.0], 0.3)
    True
    """
```
\end{lstlisting}
\caption{EBT test code generation prompt example}
\label{fig:ebt-prompt}
\end{figure}

%% file: contents/section4.tex
\section{Results}
\label{sec:results}


This section reports the results of the case study designed in Section \ref{sec:case-study-design}. We first present an overview of the experimental results, followed by detailed analysis results for RQ1 and RQ2.

\input{tables/tab-experiment-comparison}


\subsection{RQ1: How effective are PBT and EBT respectively in detecting edge cases in LLM-generated code?}


For RQ1, we evaluate how effective PBT and EBT are respectively in detecting edge cases in LLM-generated code. Table \ref{tab:experiment_comparison} shows the evaluation results of PBT and EBT for 16 edge cases. Overall, both methods detect bugs in 11 out of 16 cases (marked with \checkmark), while failing to detect bugs in 5 cases. However, the cases where each method succeeds or fails in detection differ, revealing the complementary characteristics of both methods. From the results in Table \ref{tab:experiment_comparison}, the following detection rates are obtained:


\begin{itemize}
    \item \textbf{Overall detection rate}: Both methods detected bugs in 11 out of 16 cases (68.75\%)
    \item \textbf{Method-specific success patterns}:
    \begin{itemize}
        \item PBT alone succeeded: 2 cases (HumanEval/55, 95)
        \item EBT alone succeeded: 2 cases (HumanEval/140, 150)
        \item Both methods succeeded: 9 cases
        \item Both methods failed: 3 cases (HumanEval/49, 123, 127)
    \end{itemize}
    \item \textbf{Combined detection rate}: At least one method detected bugs in 13 cases (81.25\%)
\end{itemize}


These results show that while each method has a detection rate of 68.75\% individually, combining both methods improves this to 81.25\%. This 12.5-point improvement suggests that both methods have strengths for different types of edge cases.


While computing these detection rates, we exclude at least one wrong test case in 5 out of 16 cases for PBT and in 12 out of 16 cases for EBT. Specifically, in PBT, we exclude 7 out of 146 test cases in total, while in EBT, we exclude 48 out of 468 test cases across 16 cases. In PBT, 2 of the 7 exclusions result from test code errors such as nonexistent function calls or missing arguments, while 5 exclusions result from logical errors in the test cases. Overall, PBT requires fewer corrections than EBT. We believe that rule-based automation can reduce test code errors such as missing arguments.


We also apply McNemar's test to the paired outcomes across the 16 tasks to evaluate whether there is a significant difference between Property-based Testing (PBT) and Example-based Testing (EBT). The discordant pairs are 2 cases where only PBT succeeded and 2 cases where only EBT succeeded. As a result, we find no statistically significant difference (p = 1.00, $\alpha$ = 0.05). For reference, the continuity-corrected statistic is $\chi^2 = 0.25$, which is likewise not significant. These results support the interpretation that the primary benefit lies in complementary coverage rather than superiority of one approach.



\subsection{RQ2: What types of edge cases do each method fail to detect, and what are the underlying causes?}


For RQ2, we analyze the types of edge cases where each method fails to detect bugs and their underlying causes. Below, we examine in detail the cases where each method shows superiority and cases where both methods fail.


\subsubsection{PBT-Advantageous Cases}


Specific cases where PBT shows superiority include HumanEval/55 and HumanEval/95. In HumanEval/55, as shown in Figure \ref{fig:pbt_55}, PBT sets the maximum value to \texttt{max\_value=100} in the \texttt{@given} decorator, automatically generating a wide range of inputs. In contrast, EBT only tests up to $n=35$. This approximately 2.9-fold range difference allows PBT to detect timeout issues with large-scale inputs. In HumanEval/95, using the \texttt{st.dictionaries} strategy shown in Figure \ref{fig:pbt_95}, PBT automatically generates dictionaries with arbitrary numbers of keys and string patterns. This diverse dictionary structure detects the bug that skips checking keys beyond the second, while EBT's manually created limited test cases miss this edge case.


\subsubsection{EBT-Advantageous Cases}


Cases where EBT shows superiority include HumanEval/140 and HumanEval/150. In HumanEval/140, as shown in Figure \ref{fig:ebt_140}, EBT includes an explicit test case expecting \texttt{"\_\_"} for two trailing spaces (\texttt{"  "}). This edge-case-specific test detects the bug where only one underscore is generated for two trailing spaces. In contrast, PBT's string generation patterns fail to produce this specific string structure. In HumanEval/150, as shown in Figure \ref{fig:ebt_150}, EBT implements 10 comprehensive edge case tests including negative numbers and zero such as $n=0, -1, -7$. These tests detect the bug where the function always returns $x$ (prime case) when $n \leq 0$. In contrast, PBT sets the constraint $n \geq 1$ in all tests, missing this critical boundary condition.


\subsubsection{Limitations of Both Methods}


In HumanEval/49, HumanEval/123, and HumanEval/127, both methods miss bugs. Specifically in HumanEval/49, the boundary condition $(n=0, p=1)$ is not considered by either method. In HumanEval/123 and HumanEval/127, range limitations of very large integers (maximum 1024 in EBT, 1000 in PBT) prevent detection of precision issues and timeout problems.


\subsection{Supplementary Analysis}


\subsubsection{Execution Time Comparison}


In our experiments, the average execution time for PBT is 2.540 seconds, while EBT requires 1.046 seconds on average (with a 15-second timeout). PBT takes approximately 2.4 times longer than EBT due to its automatic generation and testing of diverse inputs. However, this time difference remains within practical limits and is acceptable considering the additional bug detection capability.


\subsection{Summary of Findings}


The results of this case study demonstrate that PBT and EBT have complementary characteristics in edge case detection. While each method's individual detection rate is 68.75\%, combining both methods improves this to 81.25\%. This 12.5-point improvement reveals the following:


\begin{itemize}
    \item \textbf{PBT strengths}: Detection of performance issues and hidden logic errors through extensive input space exploration
    \item \textbf{EBT strengths}: Explicit testing of specific boundary conditions and special patterns
    \item \textbf{Common limitations}: Restrictions on special boundary condition combinations and extremely large numerical ranges
\end{itemize}


These insights suggest the effectiveness of a hybrid approach that strategically combines both methods rather than relying on a single method for LLM-based test generation.

\input{figures/pbt55}
\input{figures/pbt95}
\input{figures/ebt140}
\input{figures/ebt150}

%% file: tables/tab-experiment-comparison.tex
\begin{table}[!t]
\renewcommand{\arraystretch}{1.3}
\caption{Comparison of PBT and EBT Results for Selected HumanEval Tasks}
\label{tab:experiment_comparison}
\centering
\begin{tabular}{l|c|c}\hline\hline
Task ID & PBT & EBT \\\hline
HumanEval/22 & \checkmark & \checkmark \\
HumanEval/44 & \checkmark & \checkmark \\
HumanEval/49 &  &  \\
HumanEval/55 & \checkmark & \\
HumanEval/63 & \checkmark & \checkmark \\
HumanEval/64 & \checkmark & \checkmark \\
HumanEval/76 & \checkmark & \checkmark \\
HumanEval/95 & \checkmark & \\
HumanEval/97 & \checkmark & \checkmark \\
HumanEval/122 & \checkmark & \checkmark \\
HumanEval/123 &  &  \\
HumanEval/124 & \checkmark & \checkmark \\
HumanEval/127 &  &  \\
HumanEval/132 & \checkmark & \checkmark \\
HumanEval/140 & & \checkmark \\
HumanEval/150 & & \checkmark \\\hline
\end{tabular}
\end{table}

%% file: figures/pbt55.tex
\begin{figure*}[!t]
\begin{lstlisting}[
    basicstyle=\small, %or \small or \footnotesize etc.
]
@given(st.integers(min_value=0, max_value=100))
def test_fib_non_negative_input(n):
    """Test that fib returns non-negative results for non-negative inputs"""
    result = fib(n)
    assert result >= 0, "Fibonacci numbers should be non-negative for non-negative inputs"
\end{lstlisting}
\caption{PBT test code of HumanEval/55 (fib) (excerpt)}
\label{fig:pbt_55}
\end{figure*}

%% file: figures/pbt95.tex
\begin{figure*}[!t]
\begin{lstlisting}[
    basicstyle=\small, %or \small or \footnotesize etc.
]
@given(st.dictionaries(st.text(alphabet='abcdefghijklmnopqrstuvwxyz', min_size=1), st.text(), min_size=1))
def test_all_lowercase_strings_returns_true(test_dict):
    """Property: Dictionary with all lowercase string keys should return True"""
    result = check_dict_case(test_dict)
    assert result == True, "Dictionary with all lowercase string keys should return True"
\end{lstlisting}
\caption{PBT test code of HumanEval/95 (check\_dict\_case) (excerpt)}
\label{fig:pbt_95}
\end{figure*}

%% file: figures/ebt140.tex
\begin{figure*}[!t]
\begin{lstlisting}[
    basicstyle=\small, %or \small or \footnotesize etc.
]
assert fix_spaces("  ") == "__", "Edge test: Two spaces only"
\end{lstlisting}
\caption{EBT test code of HumanEval/140 (fix\_spaces) (excerpt)}
\label{fig:ebt_140}
\end{figure*}

%% file: figures/ebt150.tex
\begin{figure*}[!t]
\begin{lstlisting}[
    basicstyle=\small, %or \small or \footnotesize etc.
]
assert x_or_y(1, 100, 200) == 200, "Edge Test: 1 is not prime by definition, should return y=200"
assert x_or_y(0, 50, 75) == 75, "Edge Test: 0 is not prime, should return y=75"
assert x_or_y(-1, 30, 40) == 40, "Edge Test: Negative number -1 is not prime, should return y=40"
assert x_or_y(-7, 10, 20) == 20, "Edge Test: Negative number -7 is not prime, should return y=20"
assert x_or_y(2, -5, -10) == -5, "Edge Test: 2 is prime with negative x, should return x=-5"
assert x_or_y(6, -15, -25) == -25, "Edge Test: 6 is not prime with negative y, should return y=-25"
assert x_or_y(23, 0, 1) == 0, "Edge Test: 23 is prime, should return x=0"
assert x_or_y(25, 1, 0) == 0, "Edge Test: 25 is not prime (5*5), should return y=0"
assert x_or_y(2, 5, 5) == 5, "Edge Test: Same x and y values with prime n, should return x=5"
assert x_or_y(4, 10, 10) == 10, "Edge Test: Same x and y values with non-prime n, should return y=10"
\end{lstlisting}
\caption{EBT test code of HumanEval/150 (x\_or\_y) (excerpt)}
\label{fig:ebt_150}
\end{figure*}

%% file: contents/section5.tex
\section{Discussion}
\label{sec:discussion}


In this study, we test 16 out of 21 cases that failed in HumanEval+ using LLM-generated PBT and EBT. As analyzed in detail in Section \ref{sec:results}, both methods detect bugs in 11 out of 16 cases and miss bugs in 5 cases. However, it becomes clear that each method has distinct characteristics, and the results showing PBT alone succeeding in 2 cases, EBT alone in 2 cases, both methods in 9 cases, and both failing in 3 cases demonstrate that combining both methods is most effective.


\subsection{Implications for LLM-based Test Generation}


Based on the specific numerical results obtained in this research, the following implications are derived for LLM-based test generation.


\textbf{Implementation Guidelines:} When implementing the hybrid approach, it is recommended to first conduct comprehensive testing with PBT, then supplement with specific boundary condition testing using EBT. Specifically, after testing a wide input range with over 100 automatically generated test cases using PBT, confirm typical input-output examples (about 5-10 cases) with EBT.


\textbf{Importance of Test Range Design:} The approximately 2.9-fold test range difference in HumanEval/55 (PBT: n=100 vs EBT: n=35) determined the detection results, highlighting the importance of specifying test ranges to LLMs. For performance testing in particular, it is necessary to explicitly specify sufficiently large input ranges.


\textbf{Specific Prompt Design Guidelines:} For PBT test code generation, it is effective to include specific instructions in prompts such as ``explicitly specify the range from minimum to maximum values,'' ``include edge cases (0, negative numbers, empty lists, etc.),'' and ``set the number of test cases to 100 or more.''


\textbf{Systematic Verification of Boundary Conditions:} Based on boundary conditions missed by both methods, such as $(n=0, p=1)$ in HumanEval/49 and negative numbers in HumanEval/150, mechanisms for systematically verifying special value combinations are needed. The introduction of automatic generation systems for all combinations including boundary values such as 0, negative numbers, maximum values, and minimum values is recommended.


\textbf{Application Strategy in Real Development Scenarios:} In actual development scenarios, a testing strategy is conceivable where PBT is prioritized for pre-release testing where quality assurance is critical, and EBT is used to supplement when rapid feedback is needed. The PBT execution time shown in this research (approximately 2.4 times longer) remains within practical limits and represents an acceptable cost considering the additional bug detection capability.


\textbf{Effectiveness of Hybrid Approach:} Among 16 cases, with PBT alone succeeding in 2 cases, EBT alone in 2 cases, both methods in 9 cases, and both failing in 3 cases, the quantitative results demonstrate that combining both methods is most effective. Specifically, by leveraging the characteristic strengths of each method, such as PBT detecting timeout issues at n=100 in HumanEval/55 and EBT detecting negative boundary conditions in HumanEval/150, it is possible to detect 12.5\% (2/16) of bugs that would be missed by individual methods.


\subsection{Threats to Validity}


\subsubsection{Validity of LLM Models and Parameters}


The LLM models and parameter settings used in this research are merely examples. When different models, versions, or parameter settings are used, the quality and characteristics of generated test code may change, potentially affecting evaluation results. To conduct a more comprehensive evaluation, it is necessary to perform comparisons targeting multiple LLM models.


\subsubsection{Validity of Properties}


In this research, we generate PBT test code by directly providing natural language specifications to LLMs. However, if the property definitions contained in the test code are inappropriate, the effectiveness of PBT may decrease.
Currently, property quality assurance depends on prompt design and subsequent human filtering, as no automated guardrails exist to verify the quality of generated properties. Developing such guardrails to ensure the reliability and validity of generated properties remains an important direction for future work.
In the future, approaches that separate property generation from test code generation, and the establishment of criteria for evaluating the validity of properties themselves, will be important for improving the quality of generated properties.

%% file: contents/section6.tex
\section{Conclusion}
\label{sec:conclusion}


This research compares and evaluates the effectiveness of Property-based Testing (PBT) and Example-based Testing (EBT) for detecting edge cases in LLM-generated code. Through an experimental study of 16 cases, these methods are observed to have complementary characteristics: PBT excels in detecting performance issues and hidden logic errors through extensive input space exploration, while EBT excels in explicit verification of specific boundary conditions and special patterns. By combining both methods, additional bugs that would be missed by individual methods can be detected, suggesting the potential effectiveness of a hybrid approach. However, systematic limitations in special boundary condition combinations and extremely large numerical ranges remain as challenges, requiring broader verification in the future.


Future research should focus on verifying the generalizability of these findings through evaluation with multiple LLM models and validation with larger-scale datasets. Additionally, by improving property generation methods and developing automated test generation systems that leverage the characteristics of both approaches, higher-quality code generation using LLMs is expected to be realized. The insights from this research provide initial guidance for test generation strategies in LLM-based code generation contexts.